**Title:** Narrowband parallel coherent LiDAR with frequency interleaving

**Authors:**
Long Wang[1], Liang Hu[1], Wenhai Jiao[3], Yaxin Shang[1], Jianping Chen[1,2], Guiling Wu[1,2,*]

**Affiliations:**
[1]State Key Laboratory of Advanced Optical Communication Systems and Networks, Department of Electronic Engineering, Shanghai Jiao Tong University, Shanghai, China.
[2]Shanghai Key Laboratory of Navigation and Location-Based Services, Shanghai, China.
[3]Beijing Institute of Tracking and Telecommunication Technology, Beijing, China.

*Corresponding author. Email: wuguiling@sjtu.edu.cn

**Abstract**
The high demand for 3D imaging in intelligent robotics is motivating the advances of coherent LiDARs towards high performances with low complexity/cost. However, the current coherent LiDARs suffer from the tight coupling between the high ranging-imaging performance and the high complexity/cost. Herein, we propose a narrowband parallel coherent LiDAR with frequency-interleaving architecture. The LiDAR architecture utilizes narrowband signals for ranging, and interleaves multi-channel sparse and narrowband signals in frequency domain at the receiving end to significantly reduce the required bandwidth and the number of detection branches, facilitating massive parallelization with low system complexity/cost. In experiments, a ranging precision of 0.49 mm that approaches the shot noise limit, and a power sensitivity of -95 dBm (~9 photons) are achieved. Parallel 3D imaging with an equivalent imaging rate of 10 Mpixel/s and a 2 cm ranging precision is also demonstrated using only two 150 MHz receiving branches. With these desirable properties, this new LiDAR opens an avenue for the LiDAR ecosystem.

**MAIN TEXT**

**Introduction**

Three-dimensional (3D) imaging is in high demand with the development of intelligent technologies, which include autonomous driving (1, 2), mobile robotics (3, 4), virtual/augmented reality (5), digital twin (6), and other applications (7, 8). Light detection and ranging (LiDAR) techniques have attracted considerable attention as one of the most important 3D imaging tools owing to their advantages in ranging precision, angular resolution, and reliability under weak illumination conditions (9). The development of LiDAR is being driven by the need for high ranging and high imaging performances with low complexity and cost, which can promote the improvement of intelligent robotics in route planning (10, 11), profile recognition and object grasping (12). Pulsed LiDARs with centimetre-level precision have been successfully used in advanced driver-assistance systems (13,14). However, this type of LiDAR is prone to interference by crosstalk signals from adjacent channels or other external light sources (15,16). Besides, the high bandwidth requirements for the modulation, reception, and processing of shorter pulses limit its attainable precision. Coherent LiDARs can prevent interference by coherent detection and are compatible with photonic integration chip (PIC) technologies for all solid states. Currently, coherent frequency-modulated continuous-wave (FMCW) LiDAR and chaotic LiDAR have been extensively studied and are leading advances in high-imaging-rate parallelization (17-20) and PIC technologies (21-26). The two LiDARs inherently achieve high-ranging precision by taking advantage of broadband signals (9). However, broadband signals also impose high requirements on signal generation, modulation, reception, and processing, which further affects the LiDAR performance, complexity, cost, flexibility, and optoelectronic integration, thus hindering their widespread application in existing and emerging fields.

For chaotic LiDARs, the random time-varying features and high instantaneous bandwidth of the chaotic signals require high-bandwidth photodetectors (PDs) and analogue-to-digital converters (ADCs) in both the receiving and reference branches (several GHz in general) (27). The massive number of broadband optoelectronic receiving and reference branches (20) is a severe impediment to realizing high-imaging-rate parallelization because of the dramatically increased system cost, complexity, and difficulties in optoelectronic integration. FMCW LiDAR reduces the receiving bandwidth to several hundred MHz by delayed homodyne detection. However, the broadband, fast and linear frequency sweeping of the laser required for high ranging precision and high imaging rate causes several issues and/or additional processing, including degradation of the laser linewidth (28-30) and correction of linearization (16). Moreover, in the parallel pattern for a high imaging rate, to distinguish the signals of different channels, massive receiving branches or a single receiving branch with several GHz bandwidths are still required (17-19), which is still hard to be afforded in many scenarios. Overall, the total bandwidth requirement of parallel coherent detection, i.e. bandwidth required in each detection branch times the number of detection branches, has gone beyond 10 GHz and even 400 GHz and has become one of the main bottlenecks for the parallel coherent LiDAR.

Compared to broadband signals and systems, narrowband signals and systems have several characteristics that can be preferred for LiDAR systems to improve the measurement performance, reduce the system complexity and cost, and extend their application potential. First, narrowband signals are easier to generate, receive, and process, which can significantly reduce system complexity, cost, and performance limitations induced by broadband signals. Second, the phase delay of narrowband signals can be measured with high precision and is almost unrelated to the frequency of the signal (31). Based on this, fine ranging can be achieved and improved by easily increasing the frequency of the signal rather than its bandwidth. Moreover, a narrowband signal occupies a small bandwidth, is frequency determinacy and easily distinguishable from other signals in the frequency domain. These features provide a physical basis for massive parallelization with a small total receiving bandwidth. Finally, the parameters of narrowband signals, including the frequency, amplitude, and phase, are relatively easy to tune in the electric domain, enabling LiDARs with high flexibility for various applications. A sine modulation waveform is adopted by amplitude-modulated continuous-wave (AMCW) LiDARs to achieve millimetre precision (31-34). However, adopting a pure sine wave introduces a trade-off between the unambiguous range and ranging precision. Moreover, the current parallel AMCW LiDARs (e.g. the flash LiDAR) adopts the incoherent direct detection, generally suitable for short-distance 3D imaging (9,31).

In this study, we propose a narrowband parallel coherent LiDAR with frequency interleaving architecture. In the LiDAR architecture, narrowband signals are adopted for ranging, multi-channel sparse and narrowband signals are interleaved in frequency domain to compress receiving bandwidth, which can significantly reduce the required bandwidth and the number of detection branches and enbales massive parallelization while lowering system complexity and cost. For the proposed architecture, a narrowband phase-hopping subcarrier (PhS) signal is designed, which is a single-tone subcarrier signal with a $\pi$-phase hopping in each measurement duration. The unambiguous long-range and high-precision distance measurement is performed by demodulating the fine phase delay of the subcarrier and the coarse time delay of the $\pi$-phase hopping point. Based on the designed PhS signal and basic ranging principle, a narrowband frequency-interleaved parallel coherent LiDAR, phase-hopping subcarrier modulation continuous wave (PhSMCW) LiDAR, is demonstrated by modulating the PhS signal on an optical comb. Using the narrowband feature of the PhS signal, multichannel beat signals are interleaved in the frequency domain and received by one receiving branch with significantly reduced bandwidths. A joint phase and amplitude demodulation in the frequency domain (JPAD-F) algorithm is proposed to extract the ranging distances and velocities of different channels from the frequency-interleaved signals with high receiving sensitivity and negligible inter-channel crosstalk. Using a 20 MHz PhS signal, a ranging precision of 0.49 mm is achieved, which approaches the shot noise limit and could be improved to the micrometre level by simply increasing the PhS signal frequency. Using a 6-MHz PhS signal and two receiving branches with only 150 MHz bandwidths, parallel ranging with an imaging rate of 10 Mpixel/s and 2-cm ranging precision is demonstrated. Benefiting from coherent detection and frequency-domain processing of narrowband signals, an unprecedented power sensitivity of -95 dBm (~9 photons) is achieved, leading to a potential ranging distance over 300 m.

**Results**

**Principle**

Figure 1a depicts the proposed parallel PhSMCW LiDAR. Two optical frequency combs (OFCs) with comb line spacing of $f_R$ and $f_R + \Delta f_R$ are generated, serving as the signal comb (OFC1) and the local oscillator (LO) comb (OFC2), respectively. The OFCs can be generated by the micro-resonators or the electro-optic modulators, as long as the frequencies of comb teeth are locked mutually. The designed narrowband PhS signal (upper left of Fig. 1a) is modulated on the phase of all the optical carriers of OFC1 using a phase modulator. The modulated OFC1 is dispersed in different spatial directions using a multibeam scanner. The generation of a narrowband PhS signal and its external modulation on the OFC can eliminate the demand for high-speed and high-cost electrical and optoelectronic devices while avoiding degradation of the laser linewidth.

On the receiving side, the echoes of OFC1 are coherently detected using OFC2 at a single PD with a multichannel frequency-interleaved (McFI) coherent receiving scheme. The comb line spacing difference ($\Delta f_R$) between the two OFCs is reduced, making the beating components of different channels interleaved with each other in the frequency domain (Fig. 1c). Because the designed PhS signal has a narrowband at a fixed subcarrier frequency and the beat frequencies are unrelated to distance, the beat components of each channel are narrow, frequency determinacy, and sparse. Therefore, multichannel beating-frequency interleaving can be performed, and the required receiving bandwidth is approximatively equal to (SI Note 1):

$$B_{RX} = 2f_{sub} + N(2f_{mDFS} + f_c) = 2f_{sub} + N\Delta f_R \qquad (1)$$

where $f_{sub}$ is the subcarrier frequency of the PhS signal, $f_{mDFS}$ is the maximum Doppler frequency shift (DFS) related to the velocity of the target, and $f_c$ is the minimum frequency interval required to avoid crosstalk between the two adjacent beating components (SI Note 6). The McFI coherent receiving scheme enables a significant compression of the required receiving bandwidth using spare bands among the sparse and narrowband beating components. Moreover, the proportional coefficient between the required receiving bandwidth and the parallel channel number is $2f_{mDFS} + f_c$ (i.e. $\Delta f_R$) but not $2(f_{mDFS} + f_{sub})$ for the case without multi-channel frequency interleaving. Because $2f_{mDFS} + f_c$ is much less than $2(f_{mDFS} + f_{sub})$, the increase in the required receiving bandwidth with the parallel channel number is significantly reduced, facilitating many more parallelised channels. Consequently, only one (or two, see the experimental results) set of low-bandwidth PD and low-sampling-rate ADC is required for massive parallel receiving, which can enable significant savings in hardware complexity and cost and is more beneficial for optoelectronic integration. The low sampling rate also reduces the amount of data and computational cost.

A JPAD-F algorithm is proposed to obtain the distance and velocity of the target from the McFI beat signal by demodulating the time delay of the PhS signals and DFS of the optical carriers for different channels. Specifically, Fast Fourier transforms (FFT) are first performed on the multichannel beat signal (Fig. 1b) to obtain the multichannel beat spectrum (Fig. 1c), and the beat frequencies (including the baseband and sidebands) belonging to each channel are identified based on the frequency and amplitude relationships (see Methods). Subsequently, the time delay of the $i$th-channel PhS signal is obtained by demodulating the integral $T_{P,i}$ and fine fractional subcarrier period time delays $\Delta t_i$ (Fig. 1e), respectively. $T_{P,i}$ is determined by identifying the $\pi$-phase hopping in the PhS signal based on the amplitude variation of the beat sidebands when moving the window of FFT (SI Note 3). $\Delta t_i$ is determined by the subcarrier phase delay of the PhS signal $\Delta\varphi_i$ with $\Delta t_i = \Delta\varphi_i/(2\pi f_{sub})$. $\Delta\varphi_i$ can be obtained by measuring the phase difference of the ±1st order beat sidebands in the phase spectrum of FFT. The distance of the $i$th channel can be calculated as

$$L_i = c \times T_{P,i}/2 + c \times \Delta\varphi_i/(4\pi f_{sub}), \qquad (2)$$

where $c$ is the speed of light. The DFS of the $i$th-channel optical carrier can be obtained using $f_{i,\text{DFS}} = f_{i,0} - i\Delta f_R$, where $f_{i,0}$ is the identified beat baseband frequency, and $i\Delta f_R$ represents the predetermined beat baseband frequency without DFS. The velocity of the $i$th channel can then be calculated with $V_i = cf_{i,\text{DFS}}/v_{i,0}$, where $v_{i,0}$ is the $i$th-channel optical carrier frequency.

The phase measurement of narrowband signals in the frequency domain is used for fine distance measurements in the proposed LiDAR. The phase of narrowband signals can be measured with high precision and is independent of the signal frequency. Therefore, high-precision ranging could be achieved by increasing the subcarrier frequency of the PhS signal. The ranging precision limited by the shot noise (SI Note 2) is as follows:

$$\sigma_L = \frac{c\sqrt{\eta h v}}{16\pi\eta J_1\left(2\pi \frac{V_{sub}}{V_\pi}\right)} \times \frac{1}{f_{sub}\sqrt{T_S P_S}}, \qquad (3)$$

where $\sigma_L$ is the ranging precision evaluated using the standard deviation of multiple measurements, $T_S$ is the signal duration in a single measurement, $P_S$ is the signal power, and the definition of other parameters is given in SI Note 2. Although a larger receiving bandwidth is required for a higher subcarrier frequency of the PhS signal, the increase in the receiving bandwidth caused by the increase in subcarrier frequency is not multiplied by the channel number according to Eq.(1). This means that the restriction between the ranging precision and parallel channel number can be largely broken, and high ranging precision and massive parallelization can be achieved simultaneously with low receiving bandwidth.

**Fig. 1: Principle of the proposed parallel PhSMCW LiDAR. a**, System structure of the parallel PhSMCW LiDAR. OC: optical coupler, OFCG: optical frequency comb generator, PM: electro-optic phase modulator, circ.: optical circulator, DAC: analog-to-digital converter, DSP: digital signal processor. **b**, Schematic waveform of multichannel beat signal. **c**, Schematic frequency-interleaved spectrum of multichannel beat signal. Dashed lines represent the beat baseband of each channel, and the solid lines are the beat sidebands. **d**, Separation of the beat components belonging to different channels. **e**, Diagram of different time delays of the PhS signals in different channels. **f**, Schematic reconstructed 3D image of the target.

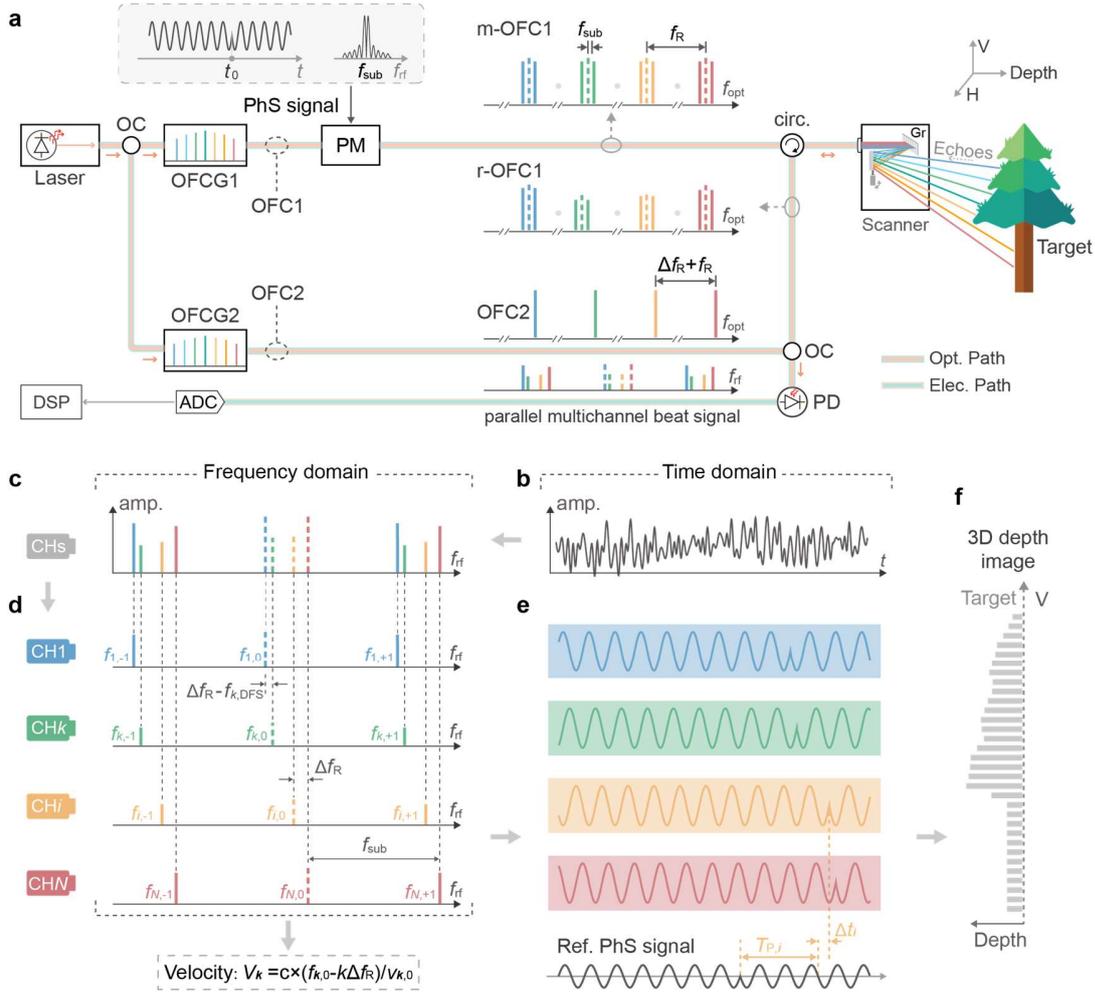

### Ranging performance evaluation

In the experiment, we first evaluated the ranging performance offered by the narrowband PhS signal and JPAD-F algorithm under a single channel. The experimental setup is shown in Fig. 2a. A continuous-wave laser was split into two parts: one served as the optical signal carrier and was modulated by the PhS signal, while the other was frequency shifted by an acousto-optic modulator (AOM) as the LO light. The PhS signal was generated using an arbitrary function generator (AFG). An optical variable delay line (OVDL) and variable optical attenuator (VOA) were used to simulate space length variation and signal attenuation. The beat signal was sampled using an oscilloscope and digitally processed offline. Figure 2b shows the frequency-domain amplitude spectrum of the beat signal at the optical signal power of -85 dBm. The beat sidebands generated by the PhS signal are distributed symmetrically on both sides of the beat baseband. In signal processing, the amplitude and phase variation of the ±1st beat sidebands are used in the JPAD-F algorithm to demodulate the length variation.

Figure 2c presents the results of length variation from 1.5 mm to 20.854 m when the parameters of the PhS signal are set at $f_{sub} = 6$ MHz and $T_S = 10$ μs. For each length, 100 successive measurements were performed. The ranging accuracy (evaluated by the mean value of 100 successive measurement errors) is better than 0.45 mm and the ranging precision (evaluated by the standard deviation ($\sigma_L$) of measurement errors) is better than 1.35 mm (Fig. 2c). When the $f_{sub}$ is increased to 20 MHz, the ranging precision is improved to a sub-millimetre level with $\sigma_L = 0.49$ mm (Fig. 2e). Moreover, the ambiguous period (7.5 m for $f_{sub} = 20$ MHz, determined by $c/2f_{sub}$) in the length of 20.854 m is accurately demodulated by identifying the π-phase hopping point (SI Note 3), and a sub-millimetre level precision is maintained. To the best of our knowledge, our ranging method represents one of the simplest ways to achieve such high precision, releasing the high-bandwidth requirement and complex operation of FMCW and chaotic LiDARs.

The ranging precision ($\sigma_L$) under different optical signal power ($P_S$) was measured and is shown in Fig. 2e and 2f. $\sigma_L$ varies with a tendency of $(P_S)^{-0.5}$, and the ranging precision approaches the theoretical shot-noise limit (SI Note 2) shown by the dashed lines. This slight deviation is primarily caused by the electrical floor noise of the PD. When the signal power is -85 dBm (SNR is 12.2 dB according to Fig. 2b), the ranging precision is about 30 cm ($f_{sub}$ = 6 MHz) and 9 cm ($f_{sub}$ = 20 MHz), respectively. Thus, the system is predicted to be capable to detect a −95 dBm (~ 9 photons) optical signal with a 2.2 dB SNR and ranging precision of 30 cm ($f_{sub}$ = 20 MHz). This weak power detection capability can support a 300 m ranging distance (SI Note 8), fulfilling the typical demand for autonomous driving.

Figure 2g shows the measurements of the simultaneous distance and speed variations. The Doppler frequency shift is simulated by adjusting the driving frequency of the AOM. The step of the distance variation is 1.5 mm, and the speed is varied from 0 m/s to 1 m/s and up to 10 m/s. $\sigma_L$ at each length is less than 0.61 mm and the standard deviation of speed measurement error is approximately 0.06 mm/s, verifying the ability of simultaneous high-precision distance and speed measurement.

**Fig. 2: Ranging performance evaluation of the PhSMCW LiDAR under single channel.** **a**, Setup for ranging performance evaluation. CWL: continuous-wave laser. AFG: arbitrary function generator, OSC: oscilloscope. The AOM is driven by a 200 MHz microwave signal. **b**, Frequency spectrum of the beat signal with a LO power of -10 dBm and a signal power of -85 dBm. **c** and **e**, Measured relative length when the OVDL is tuned with setting $f_{sub}$ = 6 MHz, $T_S$ = 10 μs and $f_{sub}$ = 20 MHz, $T_S$ = 10 μs, respectively. **d** and **f**, Variation of the ranging precision with the optical signal power. The dashed lines are the precision under the shot-noise limit calculated by Eq. (3) and the parameters in SI Note 2. **g**, Demodulated results of simultaneous distance and speed variation.

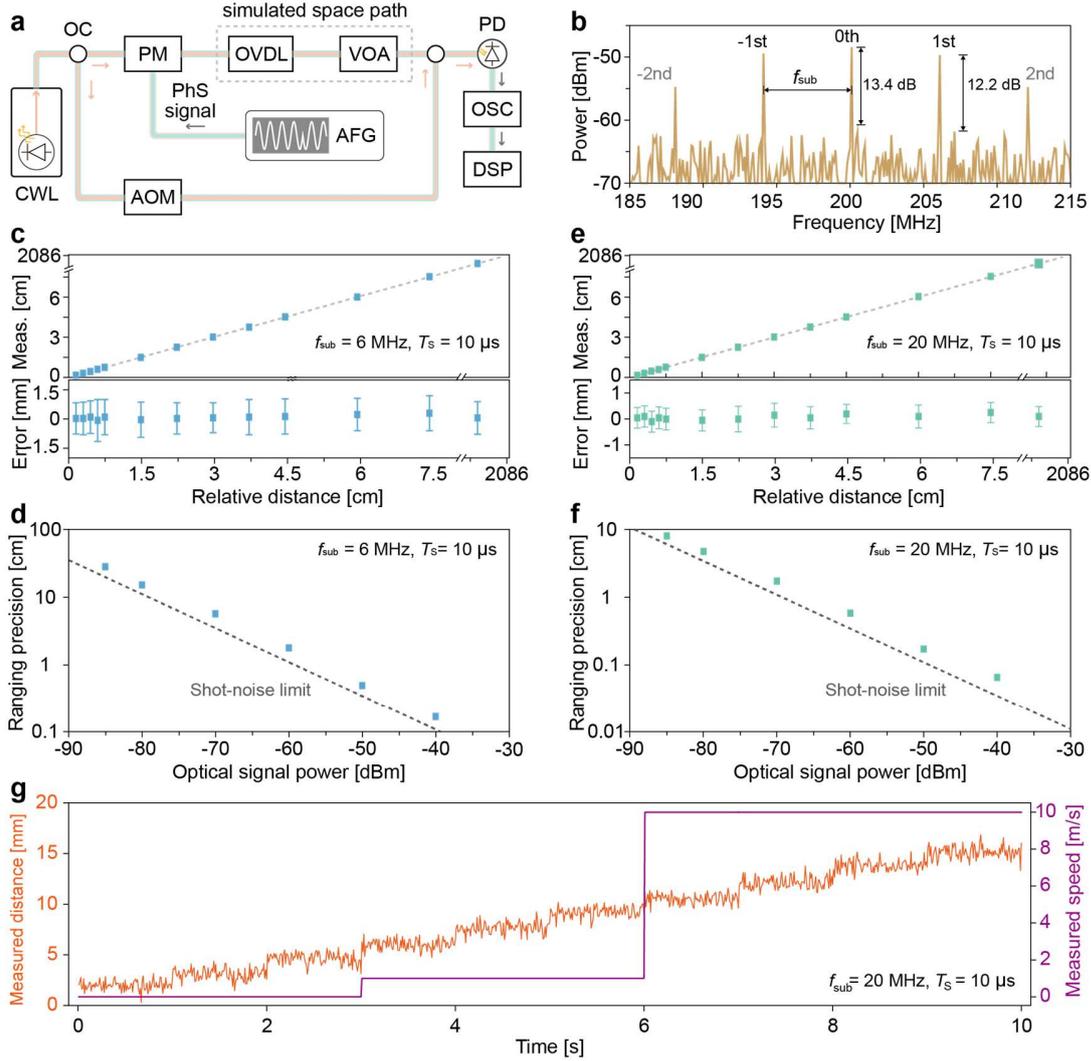

## Parallel distance and velocity measurements

We further evaluated the performance of the PhSMCW LiDAR under parallel version. Figure 3a depicts the setup. Two electro-optic comb generators were used to convert the continuous-wave seed laser into a signal comb and an LO comb. Figure 3b presents the optical spectra of the two combs. $f_R$ of the LO comb was set to 10 GHz, and that of the signal comb could be adjusted when varying the $f_{sub}$ of the PhS signal. To make full use of the spectral resource, the positive and negative comb lines (±1st to ±10th, corresponding to 20 parallel channels) were used simultaneously, and a waveshaper was used to separate the positive and negative comb lines of the LO comb to enable a separated detection of the positive and negative channels.

Parameters $f_{sub}$, $T_S$, and $\Delta f_R$ were first set to 6 MHz, 2 μs, and 15 MHz, respectively, which results in a total beat bandwidth of 150 MHz and achieves an equivalent imaging rate ($N/T_S$) of as high as 10 Mpixel/s. Figure 3c shows the partial beat spectrum of the positive channels, highlighting three adjacent channels in which the beat components of the different channels are interleaved with each other. To verify that the system is free of interchannel crosstalk, we adjusted the OVDL from 0.75 to 7.5 cm. The length variation demodulated by the three adjacent channels is shown in Fig. 3d. At each relative length, the ranging accuracy is better than 0.57 cm, and the ranging precision is about 1.21 cm, presenting no evident difference with the case without channel interleaving (SI Note 5). Figure 3e shows $\sigma_L$ of the 20 channels, which are below 2 cm. The difference of ranging precision between different channels is mainly caused by the power fluctuation of the different comb lines. Then, $f_{\text{sub}}$ and $T_S$ were increased to 20 MHz and 10 μs, respectively, and $\Delta f_R$ was adjusted to 23 MHz correspondingly. Using these parameters, an equivalent imaging rate of 2 Mpixel/s is achieved, and the total signal-beat bandwidth is 250 MHz. The ranging accuracy is 2.5 mm, and the ranging precision is improved to 3 mm (Fig. 3g). Figure 3h shows $\sigma_L$ of the 20 channels, which are below 4 mm, indicating that the increase in $f_{\text{sub}}$ and $T_S$ can also improve the ranging precision in the parallel version. As discussed in Eq. (1), the proposed McFI coherent receiving scheme enables the receiving bandwidth to increase slowly with the number of parallel channels. Thus, an imaging rate much higher than 2 Mpixel/s can be easily realised using optical comb sources with massive carriers, which can achieve millimetre precision without requiring a high receiving bandwidth.

**Fig. 3: Performance evaluation of the parallel PhSMCW LiDAR.** **a**, Setup for performance evaluation of the parallel coherent PhSMCW LiDAR system. EOCG: electro-optic comb generator, WS: waveshaper. **b**, Optical spectra of the generated two OFCs. The optical teeth at the left side of the seed laser wavelength are used as the negative channels, while the right ones are the positive channels. **c** and **f**, Partial frequency spectrum of the beat signal when setting $f_{\text{sub}} = 6$ MHz, $\Delta f_R = 15$ MHz and $f_{\text{sub}} = 20$ MHz, $\Delta f_R = 23$ MHz, respectively. **d** and **g**, Measured relative length when the OVDL is tuned. **e** and **h**, Ranging precision achieved by the 20 parallel channels when setting $f_{\text{sub}} = 6$ MHz, $T_S = 2$ μs and $f_{\text{sub}} = 20$ MHz, $T_S = 10$ μs, respectively. For the results of $f_{\text{sub}} = 6$ MHz in **d** and **e**, the maximum and minimum optical power of the comb lines is about -47.5 dBm and -53.0 dBm, respectively. For the results of $f_{\text{sub}} = 20$ MHz in **g** and **h**, the powers are about -44.0 dBm and -50.1 dBm, respectively.

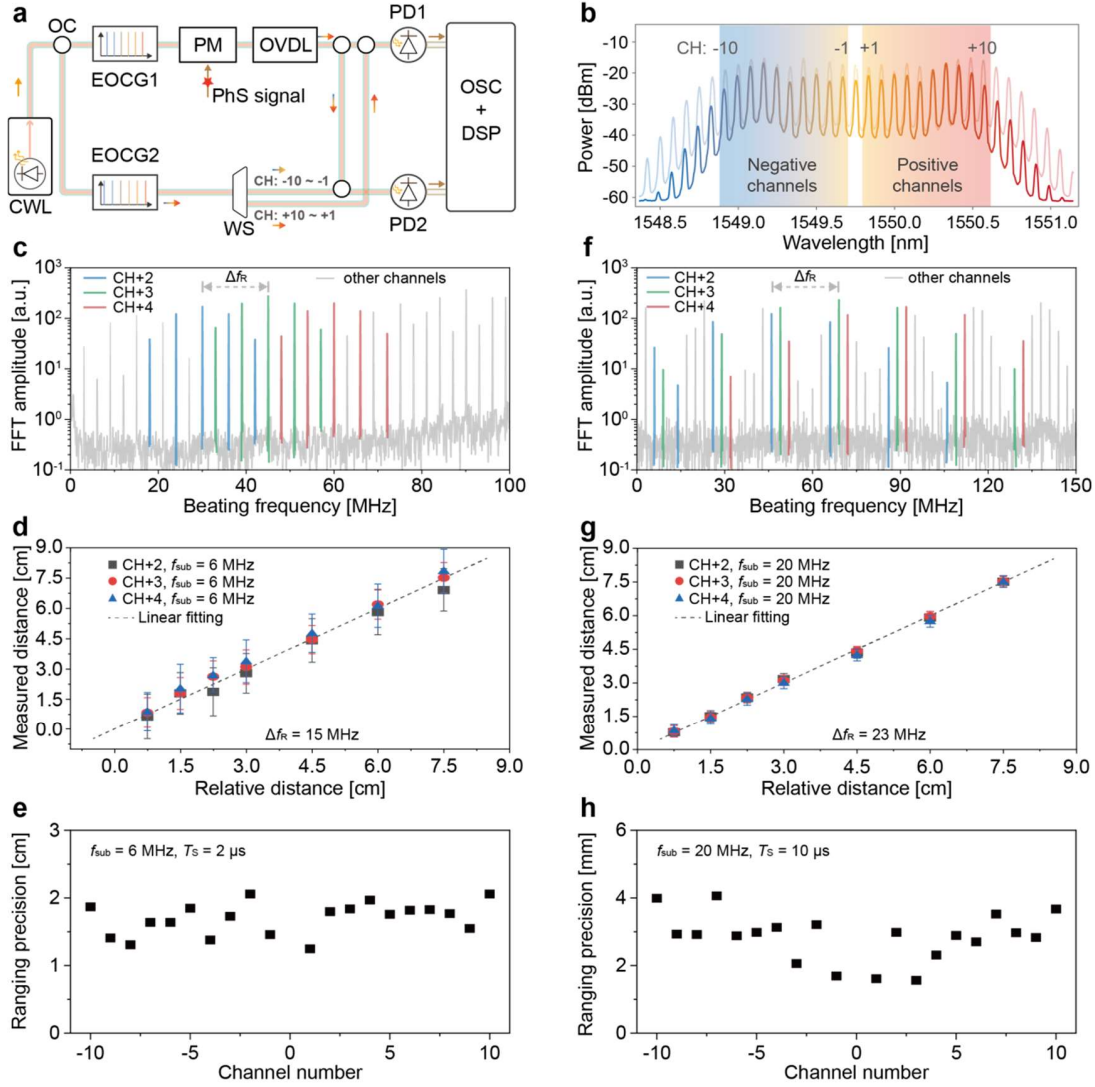

Next, we discuss the hardware requirements of the proposed parallel PhSMCW LiDAR system. (1) The dual-comb detection architecture enables the simultaneous detection of multiple channels in a single PD. When using OFCs with symmetrical spectra, as in our experimentally demonstrated system, only two PDs and ADCs are required to distinguish the positive and negative channels. (2) For the required detection bandwidth given in Eq. (1), the DFS does not exist when sensing static objects. Thus, $\Delta f_R$ can be equal to $f_c$ (approximately 3 MHz, SI Note 6) to avoid the crosstalk between the different beat components. When used in a dynamic scenario, such as in an autonomous vehicle with a maximum speed of 100 km/h, the required detection bandwidth is still only 400 MHz for 20 parallel channels, which can be easily obtained from commercial low-cost detectors. (3) For signal sampling, because the ranging precision presents no evident dependence on the sampling rate (SI Note 7), a low-rate ADC can be adopted as long as Nyquist's law is satisfied. Overall, with a reduction in hardware requirements, the proposed parallel PhSMCW LiDAR system has distinct advantages in terms of cost, complexity, flexibility, and optoelectronic integration, presenting high ranging and imaging performance.

The 3D imaging capability of the proposed parallel PhSMCW LiDAR was further explored using the setup shown in Fig. 4a. The values of $f_{sub}$, $T_S$, and $\Delta f_R$ are 6 MHz, 2 μs, and 15 MHz respectively. A bistatic system was employed, and Target 1 was formed by two sheets of reflective tape with a spacing of approximately 7 cm and a spatial distance of 0.5 m to the optical fiber collimator. Figure 4b and 4c show the reconstructed 3D cloud of points and histograms of the ranging results measured using four different probing channels, respectively. The maximum ranging error in a single measurement is < 3 cm, enabling a clear distinction between the back and front planes. The rapid-speed measurement ability was tested using a plastic plate (Target 2 in Fig. 4a). Initially, the beat baseband frequency of the $i$th channel is equal to $i\Delta f_R$. After applying an initial force, the plate generated a damped vibration and introduced DFS at the beat baseband frequency (Fig. 4d). Figure 4d presents the vibration process demodulated from the +5 and +6 channels, where the maximum vibration speed is 0.38 m/s and the duration is approximately 3.5 s. Notably, a speed below 1 mm/s is measured, further verifying the superiority of coherent LiDAR in high-precision velocity measurements, which could be used for accurate moving object recognition in autonomous vehicles, drones, and robotics. Contrarily, incoherent LiDARs rely on twice the distance measured with a certain time interval to determine the object's velocity, which makes it difficult to achieve mm/s precision owing to centimetre-level ranging precision.

**Fig. 4: Parallel 3D imaging and velocity measurement. a**, Experimental setup. WS: waveshaper, TX: transmitter, RX: receiver. WDM: wavelength division multiplexer. **b**, Reconstructed 3D image of two sheets of reflective tape. At each pixel, 1000 successive measurements were conducted, and the presented cloud of points are the averaged results. **c**, Histogram of the 1000 measurements by the adjacent ±5 and ±6 channels. **d**, Spatial velocity measurement ($V$) for an oscillating plate. The left are the frequency spectra of the beat baseband generated by the +5 and +6 channels, in which the green curves are the spectra with $V = 0$ m/s and the red curves are $V = 0.25$ m/s. The right are the vibration process measured by the +5 and +6 channels.

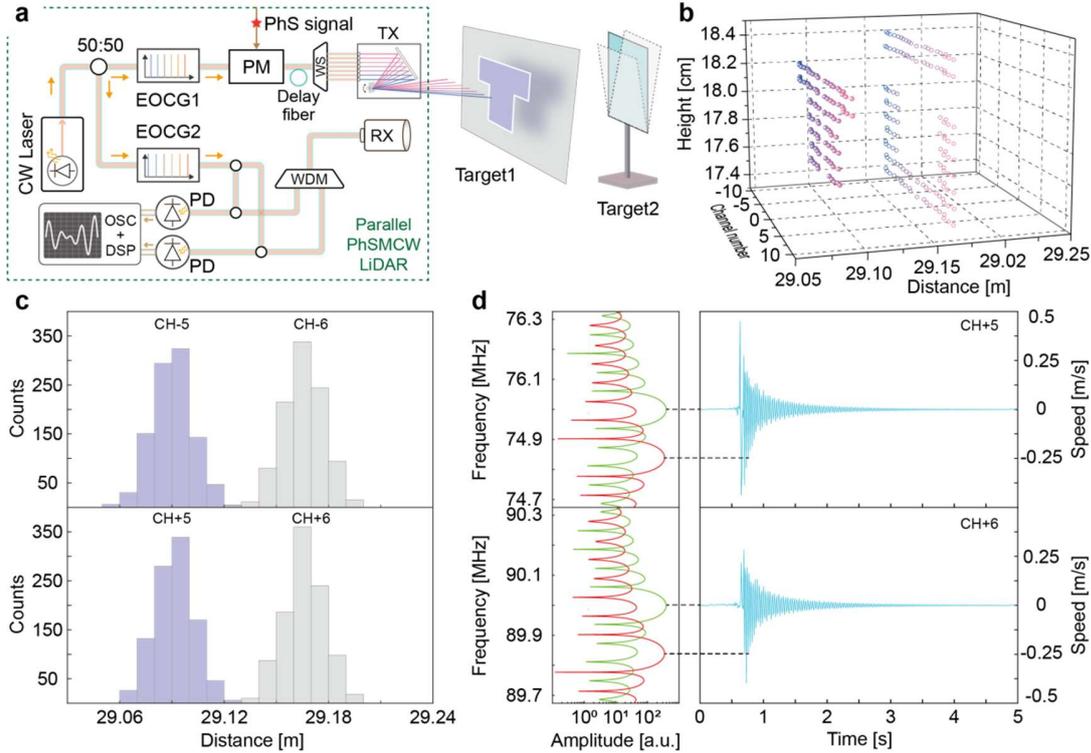

## Discussion

In summary, we demonstrated a narrowband parallel coherent PhSMCW LiDAR with advantages in 3D imaging performance, complexity, cost, and flexibility. Using the designed PhS signal and McFI coherent reception, the generation, modulation, and reception of signals are significantly simplified in the required bandwidth and receiving branch number while achieving unambiguous and high-precision parallel ranging. By avoiding the implementation difficulties and performance limitations related to broadband signals in previous parallel coherent LiDARs, PhSMCW LiDAR has the potential to achieve massive parallelisation and state-of-the-art ranging performances with low system complexity and cost. For example, on the receiving end, the required bandwidth and/or number of detection branches (including PDs and ADCs) can be reduced by one order of magnitude, as listed in Table 1. High ranging precision can be easily achieved and improved by simply increasing the subcarrier frequency of the PhS signal. Using a 1 GHz PhS, a ranging precision of up to 14 μm can be achieved by a 50-fold improvement of the experimental results at 20 MHz. An ultralow receiving sensitivity of up to −95 dBm can also be achieved. One receiving branch with significantly reduced bandwidth and negligible inter-channel crosstalk enables massive parallelisation with many channels, and therefore high imaging rate. According to the required minimum frequency internal $f_c$ and the results of $T_S = 2$ μs in Fig. 4, we can estimate that for the autonomous driving (assuming maximum speed is 100 km/h), with two receiving branch of 1.5 GHz, up to 64 channels can be supported, and a 3D imaging with an equivalent imaging rate of 32 Mpixel/s and centimetre-level ranging precision can be achieved.

Table 1. Comparison with the prior arts of parallel coherent LiDARs for 3D imaging. Since the receiving bandwidth in FMCW LiDAR system is related to the detection distance, we compare the hardware requirement under the same distance of 200 m.

| Method | Detection hardware requirement | | | Imaging rate | Comput. Cost (TFlops) | Ranging precision |
|---|---|---|---|---|---|---|
| | PD&ADC | $B_W$/PD (GHz) | Sa/ADC (GSa/s) | | | |
| FMCW (17) | 30&30 | ~ 0.5 | > 1 | 3 Mpixel/s | 1.6 | mm level |
| FMCW (18) | 2&2 | ~ 5.2 | > 10 | 5.6 Mpixel/s | 4.3 | cm level |
| FMCW (19) | 31&31 | ~ 7.8 | > 15 | 12 Mpixel/s | 28.3 | cm level |
| Chaos (20) | 80&80 | ~ 5 | > 2 | 4 Mpixel/s | 4.6 | cm level |
| This work | 2&2 | 0.15 | > 0.3 | 10 Mpixel/s | 0.025 | cm level |
| This work | 2&2 | 0.25 | > 0.5 | 2 Mpixel/s | 0.53 | mm level |

Moreover, the significantly reduced bandwidths and number of optical and electrical components render benefits to optoelectronic integration, further accelerating the advancements in massive parallel coherent LiDARs with respect to size, weight, power, and cost. Specifically, integrated continuous-wave (CW) light sources (35) and narrowband modulation (36-39) are easier to implement. The integrated electro-optic comb (40-42) can be used for parallel transmission. The optical phased array (21,43) and wavelength-division FPSA (SI Note 10) can serve as a multi-beam scanner with a sufficient field-of-view. Low-bandwidth PD and ADC are easier to integrate into a single receiving branch.

The proposed PhSMCW LiDAR also promises potential flexibility and anti-interference effects. The subcarrier frequency and duration of the PhS signal can be easily changed in the electrical domain using mature low-cost techniques to achieve different performances. This high flexibility enables the adaptation of LiDARs to universal applications. Interference among LiDARs can also be eliminated by assigning different subcarrier frequencies of PhS signals for different PhSMCW LiDARs, which realises a novel anti-interference method in addition to using coherent detection.

**Methods**

**Details in the experiment**

For the electro-optic comb generation, we used a narrow-linewidth CW laser as the seed light in experiments. As discussed in SI Note 11, the proposed LiDAR system has no high requirement on the laser linewidth, and the current chip-scale external cavity lasers and self-injection locking lasers can be employed. The EOCG adopted the scheme of cascaded modulators (44) that comprises an electro-optic phase modulator, an electro-optic intensity modulator, a RF phase shifter, a RF splitter, and two RF amplifiers. Additionally, the two EOCGs were driven by a dual-channel microwave source. For the coherent detection, two commercial balanced photodetectors were used. To avoid the power saturation, we adjusted the local light to -13 dBm.

**Frequency and phase determination in the JPAD-F algorithm**

In the JPAD-F algorithm, the beat frequencies generated by each channel need to be first identified. When detecting stationary objects, the beating frequencies of all parallel channels can be predetermined. For instance, the frequency of the beat baseband generated by the *i*th channel is equal to $i\Delta f_R$, and the frequencies of the ±1st beat sidebands are $i\Delta f_R - f_{sub}$ and $i\Delta f_R + f_{sub}$, respectively. When detecting the moving objects, the beat components will shift in the frequency spectrum due to the Doppler frequency shift. Thus, the $\Delta f_R$ needs to satisfy $\Delta f_R > 2f_{mDFS}$ to avoid the channel confusion (SI Note 1). In this case, the frequency of the *i*th-channel beat baseband lies in the range from $i\Delta f_R - f_{mDFS}$ and $i\Delta f_R + f_{mDFS}$. Based on this priori condition and the fact that frequency interval between the ±1st beat sidebands and baseband is equal to the $f_{sub}$. Then, the beat frequencies, taking the *i*th channel as an example, can be identified by the following steps:

(1) Extract all the beat frequencies ($f_{i,1}, f_{i,2}, \ldots, f_{i,n}$) from the FFT amplitude spectrum as the undetermined frequencies of the *i*th channel, whose power is higher than the noise power threshold. For instance, the power threshold in the demonstrated system can be set at -62 dBm (Fig. 2b).

(2) Select all the frequency combinations ($f_{i,x}, f_{i,y}, f_{i,z}$) from the undetermined values, whose frequency differences satisfy $f_{i,y} - f_{i,x} = f_{i,z} - f_{i,y} = f_{sub}$.

(3) According to the beat power, determine the frequency combinations ($f_{i,y}, f_{i,x}, f_{i,z}$) as the beat baseband, the -1st order beat sideband, and the +1st beat sideband of the *i*th channel, whose power difference is less than 3 dB. This is due to that we can adjust the amplitude of the PhS signal to realize an equal intensity in the carrier and ±1st sidebands, while the high-order sidebands present a relatively large power difference, as shown in Fig. 3c.

(4) From the FFT phase spectrum, extract the corresponding phase term ($\varphi_{i,y}, \varphi_{i,x}, \varphi_{i,z}$) of the frequencies ($f_{i,y}, f_{i,x}, f_{i,z}$). Then, the phase difference $\Delta\varphi_i$ of the ±1st order beat sidebands in Eq.(2) can be obtained by $\Delta\varphi_i = (\varphi_{i,z} - \varphi_{i,x})/2$, and then the $\Delta t_i = (\varphi_{i,z} - \varphi_{i,x})/(4\pi f_{sub})$.

**Supplementary Materials**

This PDF file includes:
Supplementary text
Figs. S1 to S13
Tables S1 to S5
References (9, 26, 35, 45-49)

**Acknowledgments**

We thank X. Chen and K. Xie for fruitful discussions about the experiment setup and signal processing method.

**Funding:**
National Natural Science Foundation of China 61627817
National Natural Science Foundation of China 61905143

**Author contributions:** The idea was conceived by L.W. The experiments were conceived by L.W., L.H. and G.W. The EOCGs were conducted by L.W. and L.H. Ranging experiments were conducted by L.W. with the assistance of Y.S. The results are analysed by G.W., L.W., W.J and J.P. All authors participated in writing the manuscript. The project was performed under the supervision of G.W.

**Competing interests:** The authors declare that they have no conflict of interest.

**Data and materials availability:** All data needed to evaluate the conclusions in the paper are present in the paper and/or the Supplementary Materials.